\documentclass{article} 

\usepackage{hyperref}

\usepackage[accepted]{icml2025}

\usepackage{microtype}
\usepackage{graphicx}
\usepackage{caption}
\usepackage{subcaption}
\usepackage{booktabs} 
\usepackage{algorithm}
\usepackage{algorithmic}
\usepackage{amsthm}
\usepackage{amssymb}
\usepackage{amsmath}
\usepackage{bbold}
\usepackage{siunitx}

\usepackage{amsmath,amsfonts,bm}

\def\eqref#1{equation~\ref{#1}}

\def\1{\bm{1}}

\DeclareMathAlphabet{\mathsfit}{\encodingdefault}{\sfdefault}{m}{sl}
\SetMathAlphabet{\mathsfit}{bold}{\encodingdefault}{\sfdefault}{bx}{n}

\usepackage[capitalize,noabbrev]{cleveref}
\usepackage{multirow}
\usepackage{enumitem}
\usepackage{tikz}
\usepackage{pgfplots}
\usepackage{listings}
\theoremstyle{plain}

\theoremstyle{definition}

\theoremstyle{remark}

\renewrobustcmd{\bfseries}{\fontseries{b}\selectfont}
\renewrobustcmd{\boldmath}{}
\newrobustcmd{\B}{\bfseries}

\newcommand{\ourmethod}{\textsc{\textit{BAP}}}
\newcommand{\ourmethodlong}{Bug Attention Probe}
\newcommand{\na}{\textsc{n/a}}

\icmltitlerunning{Where's the Bug? Attention Probing for Scalable Fault Localization}

\usepackage{url}
\usepackage{comment}

\begin{document}

\twocolumn 
[

\icmltitle{Where's the Bug? Attention Probing for Scalable Fault Localization}

\icmlsetsymbol{equal}{*}

\begin{icmlauthorlist}
\icmlauthor{Adam Stein}{equal,upenn}
\icmlauthor{Arthur Wayne}{equal,upenn}
\icmlauthor{Aaditya Naik}{upenn}
\icmlauthor{Mayur Naik}{upenn}
\icmlauthor{Eric Wong}{upenn}
\end{icmlauthorlist}

\icmlaffiliation{upenn}{Department of Computer Science, University of Pennsylvania, Pennsylvania, USA}

\icmlcorrespondingauthor{Adam Stein}{steinad@seas.upenn.edu}
\icmlcorrespondingauthor{Arthur Wayne\newline}{artwayne@seas.upenn.edu}

\icmlkeywords{Machine Learning, ICML}

\vskip 0.3in
 ]

\printAffiliationsAndNotice{\icmlEqualContribution} 

\begin{abstract}

Ensuring code correctness remains a challenging problem even
as large language models (LLMs) become increasingly capable at code-related tasks. While LLM-based program repair systems can propose bug fixes using only a user's bug report, their effectiveness is fundamentally limited by their ability to perform fault localization (FL), a challenging problem for both humans and LLMs.
Existing FL approaches rely on executable test cases, require training on costly and often noisy line-level annotations, or demand resource-intensive LLMs.
In this paper, we present \textit{\ourmethodlong{}} (\ourmethod{}), a method which learns state-of-the-art fault localization without any direct localization labels, outperforming traditional FL baselines and prompting of large-scale LLMs.
We evaluate our approach across a variety of code settings, including real-world Java bugs from the standard Defects4J dataset as well as seven other datasets which span a diverse set of bug types and languages. Averaged across all eight datasets, 
\ourmethod{} improves by 34.6\% top-1 accuracy
compared to the strongest baseline and 93.4\% over zero-shot prompting GPT-4o. \ourmethod{} is also significantly more efficient than prompting, outperforming large open-weight models at a small fraction of the computational cost.\footnote{\ourmethod{} is open-sourced here:\newline \url{https://github.com/adaminsky/BAP}}

\end{abstract}

\section{Introduction}

\begin{figure}[t]
    \centering    \includegraphics[width=\linewidth]{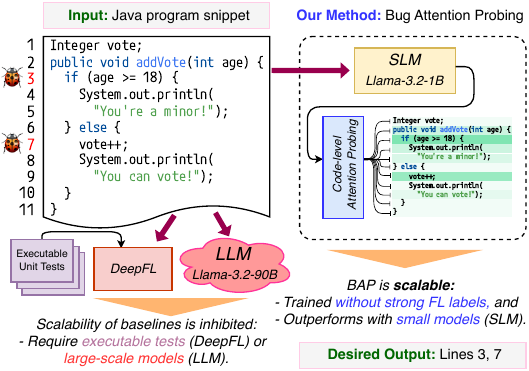}
    \caption{Comparison of our approach \textit{\ourmethodlong{}} (\ourmethod{}) with baselines DeepFL and LLM prompting on a  Java program snippet. The program has two bugs: the age condition on line 3 is reversed and line 6 throws a null pointer exception. \ourmethod{} correctly localizes both bugs. Here, our method is trained on Llama-3.2-1B, a ``small" language model (SLM), with only weak supervision \textit{i.e.} binary bug presence labels. Obtaining comparable accuracy via prompting demands a significantly more resource-intensive LLM, such as Llama-3.2-90B, or even larger. Previous approaches to fault localization like DeepFL require executable test cases before they can attempt to provide useful information.}
    \label{fig:splash}
    \vspace{-0.1in}
\end{figure}

Correctness is a fundamental desirable property of code.
Both human-written and LLM-generated code are prone to bugs \cite{karampatsis2020often, jesse2023large} including
syntax errors that prevent the execution of code, semantic mistakes that cause incorrect or unintended behaviors, and vulnerabilities that compromise security in otherwise correct code.
While there are various methods for detecting bugs (e.g. failed tests, via human bug report, program crash, etc.), identifying its root cause, or \textit{localizing} the bug, is still costly \citep{vessey1985expertise, flsurvey}.


Automated software fault localization (FL) aims to help a programmer answer the question, ``Where's the bug?'', ideally pointing to specific lines of buggy code.
The traditional FL approaches rely on executable tests to determine the buggy lines \cite{usestaticanalysis}.
Without relying on tests, FL is even more challenging since such a system must reason about what is buggy without external feedback. But recently,
supervised training of models on large, labeled datasets \citep{llmao, transferfl}, and prompting of the largest LLMs such as GPT-4o has shown promise at FL without tests \citep{wu2023large}.
On a single method context, large-scale supervised training and prompting approaches can significantly surpass the traditional techniques which need tests \citep{llmao, wu2023large}.


State-of-the-art FL methods, however, are still limited in \textit{scalability}, or the ability to leverage cheaply available supervision to reach strong performance, even with small models. This lack of scalability leads to the impracticality of many FL techniques.
For example, we show a simple buggy Java code snippet in Figure~\ref{fig:splash} which we want to run FL on, but we encounter several issues: traditional methods require executable tests and LLM prompting methods are only effective for the largest of models. On large codebases, LLMs must be called at least once per function, which quickly becomes expensive. Further, training-based approaches assume extensive amounts of strong supervision for FL, which is rarely available in practice \citep{well}.

This leads to our central question: 
\textit{How do we achieve strong fault localization performance without relying on executable tests, costly large-scale LLMs, or strong supervision?}


We answer this question by proposing the \ourmethodlong{} (\ourmethod{}), a scalable LLM probing technique for FL, scaling to use available bug related data without strong FL supervision and scaling with base model size while still achieving strong performance with small models. \ourmethod{} exhibits three desirable properties:
(1) \textbf{lightweight}, (2) \textbf{code-level localization}, and (3) localization of \textbf{multi-line bugs}. First, \textit{lightweight} refers to the limited requirement for supervision (we use an attention mechanism to learn from binary bug presence rather than fine-grained location supervision), the test-free nature, and the model size (\ourmethod{} can probe small language models (SLMs) to elicit performance significantly stronger than the underlying SLM). Second, \ourmethod{} localizes bugs in a human-interpretable manner to expressions, statements, or lines in code, even though LLMs operate on the token level. Third, \ourmethod{} localizes multi-line bugs, or multiple bugs in one method, better than existing approaches. Multi-line bugs are practically relevant since the majority of real bugs are multi-line \citep{pearson2016evaluating}.

To evaluate \ourmethod{}, we use eight diverse and widely-used fault localization benchmarks, including syntax errors, semantic mistakes, and weaknesses.
Our evaluation suite covers over 50,000 
code samples across three languages: Python, Java, and C.
This notably includes Defects4J \citep{defects4j}, the most commonly used FL dataset.


We evaluate \ourmethod{} on top of the Llama-3 family of models and compare it to state-of-the-art FL methods, including traditional test-based FL and prompting of various proprietary and open-weights LLMs.
Averaged across eight datasets, \ourmethod{} improves by 23.4\% over the strongest baselines for top-1 FL accuracy which includes a 24.2\% improvement for Defects4J, and a 50.5\% improvement on DeepFix \citep{gupta2017deepfix}.
In addition, \ourmethod{} achieves these performance improvements at over ten times greater efficiency in terms of model size and FLOPs for inference. \ourmethod{} also localizes multi-line bugs better than existing methods and continues to have stronger performance than prompting for longer code sequences.
While \ourmethod{} significantly outperforms competitive baselines, it is able to achieve 35\% top-1 FL accuracy on average over our datasets, and our evaluation highlights avenues for further advances in scalable FL.


In summary, our work makes the following contributions:
\begin{itemize}[itemsep=0pt, topsep=0pt]
    \item We propose \ourmethodlong{} (\ourmethod{}) as a general method for scalable fault localization, requiring only coarse-grained detection supervision and eschewing the need for localization labels (Section~\ref{sec:ourmethod}).
    \item
    \ourmethod{} significantly improves over state-of-the-art fault localization methods by 34.6\% top-1 accuracy on average over eight fault localization benchmarks while utilizing ten times less memory resources and FLOPs.
    \item \ourmethod{} exhibits better length generalization and can predict multi-line bugs more effectively than prompting.  We also identify areas for further advances.
\end{itemize}

\begin{figure*}
    \centering
    \includegraphics[width=\linewidth]{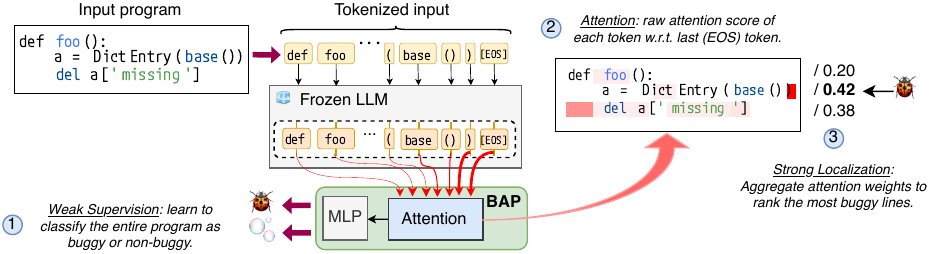}
    \caption{Illustration of \ourmethod{} as a method to elicit line-level fault localization from a frozen LLM through weak supervision. In step one, the probe is trained as a binary classifier to distinguish buggy from non-buggy code. Then in step two, we visualize the learned attention weights on the given sequence. Finally, in step three, we sum the attention weights within each line to produce a line-level ``bugginess'' score. \ourmethod{} localizes the bug to the line with the highest score, the Top-1 result.}
    \label{fig:overview}
\end{figure*}
\section{Background}
\label{sec:probing}
We introduce the fault localization problem, LLM probing, and challenges with existing FL techniques.

\subsection{Fault Localization}
\label{sec:localization}
Before defining bug localization, we first describe the problem of bug \textit{detection}, or determining if code is overall buggy or not.
A bug is a concept $b:\mathcal{P}\rightarrow \{0,1\}$ where
$\mathcal{P}$ is the space of all programs.
For a bug $b$, we construct a supervised detection dataset $\mathcal{D}_\text{Det} = \{(p, b(p)) \mid p\in \mathcal{P}\}$ consisting of programs labeled as buggy or not buggy.

Bug localization is the task of identifying the line (or lines) of code where a bug occurs. A bug localizer is a mapping $l:\mathcal{P}\rightarrow \mathbb{Z}^+$ from programs to one or more line numbers.
For a program $p \in \mathcal{P}$ split into lines $[p_0, \dots, p_k]$, where $p_i$ is the $i^\text{th}$ line of program $p$,
we define the ground truth bug localization using the notion of a counterfactual explanation: a program has a bug localized to line $i$ if modifying $p_i$ would remove the bug.
Formally, for program $p$ such that $b(p)=1$, bug $b$ is localized to line $i$ if there exists a modified line $p_i'$ such that $b(p')=0$ where $p'=[p_0, \dots, p_{i-1}, p_i', p_{i+1}, \dots, p_k]$. 

Some bugs require multi-line fixes, in which case changing multiple lines of the program would fix the bug.
We therefore assume that the ground truth localization consists of one or more lines of code. We can now define a localization dataset containing direct line-level supervision:
\begin{align*}
\mathcal{D}_\text{Loc} = &\{(p, 0, \varnothing) \mid p\in\mathcal{P} \text{ and } b(p) = 0\} \\
&\cup \{(p, 1, l(p)) \mid p\in\mathcal{P} \text{ and } b(p) = 1\}.
\end{align*}

\subsection{LLM Probing}
LLM probing involves training a classifier on top of intermediate states from a model \citep{alain2016understanding}.
Since all probing techniques are trained from a dataset of model hidden representations, we start by introducing this dataset in our setting.

For a program fragment $p$ consisting of $T$ tokens, we call the intermediate representation from LLM layer $k$, $\text{LLMRep}(p, k) = z \in\mathbb{R}^{T\times d}$ where $d$ is the hidden dimension of the LLM. The dataset we use for probing in the rest of this paper is the following:
\begin{align*}
    \mathcal{H}_{\text{Det}} = \{(\text{LLMRep}(p, k), y) : (p, y)\in\mathcal{D}_\text{Det}\}.
\end{align*}
We can also define $\mathcal{H}_\text{Loc}$ equivalently which additionally includes the ground truth line numbers with each sample, but as we discuss later, this dataset is not ideal for training.

These datasets, $\mathcal{H}_\text{Det}$ and $\mathcal{H}_\text{Loc}$, are not directly amenable to standard probing since each sample, $\text{LLMRep}(p, k)$, is a sequence of hidden representations varying in length across samples.
Producing a general-purpose fixed-length sequence representation from an autoregressive LLM is a challenging problem \cite{liu2023meaning}, so probing methods typically apply a simple pooling operation, $\text{POOL}:\mathbb{R}^{T\times d}\rightarrow \mathbb{R}^{1\times d}$, such as selecting the last token or averaging all tokens. Specific to using $\mathcal{H}_\text{Loc}$, existing work takes the approach of converting the labels $l(p)$ into a binary mask over the buggy program lines so that a model can be trained to predict such masks \citep{llmao}.

\subsection{Challenges}

We describe three main challenges with FL below.

\textbf{Need for Strong Supervision}
High-quality bug localization datasets are elusive since they require large amounts of manual effort to create. This is because a method for detecting a bug, such as a failing test case, does not immediately indicate what caused the failure. Therefore, existing real-world FL datasets are often small (e.g. Defects4J which has 395 samples \citep{defects4j}) or very noisy (e.g. ManySStuBs4J \citep{karampatsis2020often}). In our experiments, we find that methods trained on either these small or noisy datasets do not perform as well as methods which ignore such strong supervision.

\textbf{Localizing Multi-line Bugs}
The majority of real-world bugs are localized to multiple lines of code \citep{pearson2016evaluating}.
Existing methods for FL, however, mostly focus on single-line bug detection \citep{hirsch2020fault} and there are several datasets specifically catered to this setting \citep{karampatsis2020often, richter2022tssb}.



\textbf{Resource Efficiency}
The most powerful LLMs are state-of-the-art for FL tasks on a method-level context, but these models are also very resource intensive. Many of the best models are out of reach to use on the average laptop. These models are also available through APIs, but for FL settings where one may repeatedly evaluate a model on code as it changes, this can quickly become expensive. This is in sharp contrast to traditional execution-based FL methods which can run in the background of an IDE on a laptop.
\section{Attention Probing for Fault Localization}
\label{sec:ourmethod}




In this section, we introduce our approach for addressing the previous challenges.

We propose the \ourmethodlong{} (\ourmethod{}),
an LLM probing technique for performing FL which provides scalability through lightweight requirements (both in terms of training supervision and model size), interpretable code-level localization, and handling of multi-line bugs.
We take inspiration from attention probing \citep{tenney2018you, niu2022does}, a technique from the interpretability literature for studying linguistic phenomena such as the attention placed on verb tokens, but we use such an approach for code-level FL, as in localizing bugs to lines within code.

We illustrate \ourmethod{}, its process of training from bug detection data, $\mathcal{H}_\text{Det}$, and the computation of line-level bug localization from attention weights in Figure~\ref{fig:overview}.

\subsection{Why Attention Probing for Fault Localization?}
Our main motivation for using an attention mechanism is that the attention pooling operation trained for the task of bug detection encourages the probe to learn to attend to the location of bugs, without using strong FL supervision.
Intuitively, the probe attends to parts of the code which are more informative for bug detection, and we hypothesize that these locations often correspond to the location of the bugs.

For bugs which are localized to multiple lines, which is the majority of real world faults, an attention mechanism is also helpful since it operates on the token-level, making no distinction between a single and multi-line bug location.


Finally, an attention mechanism is efficient since it operates over all tokens in parallel. This means that a complete ranking of all the program lines in terms of the likelihood they contain a bug is produced by a single forward pass of attention. This is in contrast to methods such as LLM prompting which must continue to output more tokens to localize a bug to more lines.


\subsection{From Weak Supervision to Token-level Localization}
To enable learning FL from weak supervision, we
use a single layer Transformer decoder block as the architecture of \ourmethod{} to 
factorize the bug detection task into localization (attention) over tokens and detection (classification) on the sequence-level.
The input to the probe consists of a sequence of token representations from the $k$th layer of the LLM, $\text{LLMRep}(p, k) = [z_1,\dots, z_T]\in\mathbb{R}^{T\times d}$, for a program $p$ consisting of $T$ tokens. The standard multi-head attention mechanism takes $\text{LLMRep}(p, k)$ as input and outputs attention scores $a\in\mathbb{R}^{M\times T\times T}$ where $M$ is the number of attention heads, and processed token representations $v\in\mathbb{R}^{T\times d}$. We compute token-level attention scores of all tokens to the final token as $\bar{a} = \frac{1}{M}\sum_{m=1}^M a_{m,-1,:} \in\mathbb{R}^T$ where we average the attention from each head. The process of producing token-level attention from input code is also shown in lines 1-5 of Algorithm~\ref{alg:ourmethod}.


To get useful token-level attention from our model, we must train it on a downstream task. To train \ourmethod{}, we use weak supervision in the form of a bug detection dataset, $\mathcal{H}_\text{Det}$ defined in Section~\ref{sec:probing}. We pass the last token from the processed sequence $v$ from the attention through a feed-forward network (MLP) to produce a single scalar output representing the buggy/non-buggy prediction.
The model is then optimized using gradient descent with the binary cross-entropy loss.
After training the probe for bug detection, we then can examine the learned attention weights $\bar{a}$ which provide a token-level fault localization.

\subsection{From Token-level to Code-level Localization}
A major challenge with this approach is that tokens are not interpretable for programmers. Therefore, we provide a method to aggregate token-level fault localization into a code-level localization. Our approach is summarized in lines 6 to 8 of Algorithm~\ref{alg:ourmethod}. In our experiments, we focus on line-level granularity, but this method also allows us to perform statement-level and function-level localization without any significant modifications.

\begin{algorithm}[tb]
   \caption{\ourmethod{} Line-level Fault Localization}
   \label{alg:ourmethod}
\begin{algorithmic}[1]
   \STATE {\bfseries Input:} code sample $p$, layer $k$.
   \STATE Tokenize $p$ to get $p=[t_1,\dots, t_T]$.
   \STATE $z = \text{LLMRep}(p, k)$
   \STATE $v, a = \text{MHA}(z)$ where $v\in \mathbb{R}^{T\times d}$ and $a\in\mathbb{R}^{M\times T\times T}$
   \STATE $\bar{a} = \frac{1}{M}\sum_{m=1}^M a_{m,-1,:}$ \COMMENT{Average attention over all heads, for the last token}
   \STATE Group attention scores into lines: $s_1 = [\bar{a}_1, \dots, \bar{a}_i]$, $s_2 = [\bar{a}_{i+1}, \dots, \bar{a}_{i+j}], \dots$
   \STATE $l_i = \sum_{t\in s_i} t$ for all $i$ \COMMENT{Line-level attention score}
   \STATE {\bfseries Return:} $\operatorname{argsort}_{i\in[1,L]} l_i $
\end{algorithmic}
\end{algorithm}



Line 5 of Algorithm~\ref{alg:ourmethod} computes the token-level attention scores, and we call $\bar{a}_i$ the probe's attention score for the $i$th token.
To produce a line-level attention score, we sum the attention scores for all tokens in the $i$th line, $s_i = [\bar{a}_{i_1}, \bar{a}_{i_2}, \dots ]$, to produce $l_i$, the probe's attention score for the $i$th line. The computation of line-level attention scores is shown in line 7 of Algorithm~\ref{alg:ourmethod}.

Line 8 of Algorithm~\ref{alg:ourmethod} shows that the resulting line-level localization from \ourmethod{} is the ranking of input lines based on their respective attention scores. In practice, we truncate to the top-\( k \) lines out of total lines \( L \).
The line with the highest attention score is thus noted as the top-1 prediction.

\section{Experiments}
\label{sec:experiments}

We evaluate \ourmethod{} over a diverse suite of eight fault localization benchmarks. This includes Defects4J \citep{defects4j}, the most popular fault localization benchmark, as well as seven additional benchmarks covering three general bug types.
In the rest of this section, we first introduce the datasets, then the baseline methods, and finally the results of each experiment. We answer the following research questions in this section:

\begin{description}[topsep=0pt, itemsep=0pt]
    \item[RQ1] How effective is \ourmethod{} at FL in diverse scenarios?
    \item[RQ2] How does the efficiency of \ourmethod{} compare to baselines?
    \item[RQ3] Can \ourmethod{} effectively localize multi-line bugs?
    \item[RQ4] How does the generalization ability of \ourmethod{} compare to zero-shot prompting of the base model?
\end{description}

\subsection{Datasets}

We evaluate on eight datasets summarized below.

\begin{table}[t]
    \centering
    \small
    \caption{A summary of the datasets we use for our evaluation. For a further breakdown of buggy samples into categories, see our discussion in Appendix~\ref{app:datasets}.}
    \label{tab:datasets}
    \begin{tabular}{lrrrrr}
    \toprule
         & \multicolumn{2}{c}{\# Train} & \multicolumn{2}{c}{\# Test} & \\
         \cmidrule(lr){2-3} \cmidrule(lr){4-5}
         Dataset & Bug & Clean & Bug & Clean & $\mathbb{E}$[LoC]\\
         \midrule
         Defects4J v1.2.0 & 368 & 368 & 90 & 90 & 35.8\\
         Defects4J v3.0.1 & \na & \na & 437 & \na & 46.7\\
         GitHub-Py & 1323 & 1323 & 400 & 400 & 9.3\\
         GitHub-J & 1370 & 1370 & 460 & 460 & 19.3\\
         DeepFix & 1475 & 1475 & 365 & 365 & 26.3\\
         TSSB & 4085 & 3745 & 1104 & 1080 & 24.8\\
         ManySStuBs4J & 3821 & 3821 & 1093 & 1093 & 15.5\\
         Juliet-J & 4039 & 3061 & 1011 & 989 & 62.5\\
         Juliet-C & 3718 & 3697 & 966 & 939 & 44.7\\
    \bottomrule
    \end{tabular}
\end{table}

\begin{table*}[t]
    \centering
    \small
    \caption{Comparison of \ourmethod{} with existing FL methods across eight benchmarks. We evaluate line-level localization performance on a method-level context, measured by top-1 accuracy. From left-to-right: Defects4J v1.2.0, GitHub-Python, GitHub-Java, DeepFix, TSSB-3M, ManySStuBs4J, Juliet-Java, and Juliet-C. Error bars are provided in Appendix~\ref{app:results}.}
    \label{tab:defects4j}
        \begin{tabular}{lrrrrrrrr|r}
        \toprule
        \textbf{Method} & \textbf{D4J} & \textbf{GH-Py} & \textbf{GH-J} & \textbf{DeepFix} & \textbf{TSSB} & \textbf{MS4J} & \textbf{Juliet-J} & \textbf{Juliet-C} & \textbf{Avg.}\\
        \midrule
        Random & 0.144 & 0.100 & 0.134 & 0.038 & 0.069 & 0.124 & 0.025 & 0.058 & 0.087\\
        \midrule
        DeepFL & 0.144 & \na & \na & \na & \na & \na & \na & \na & 0.144 \\      
        SmartFL & 0.158 & \na & \na & \na & \na & \na & \na & \na & 0.158 \\
        TRANSFER-FL & 0.218 & \na & \na & \na & \na & \na & \na & \na & 0.218 \\
        \midrule
        CodeLlama-70B & 0.212 & 0.145 & 0.316 & 0.084 & 0.077 & 0.169 & 0.038 & 0.095 & 0.142\\
        Llama-3.3-70B & 0.269 & 0.225 & 0.272 & 0.092 & 0.114 & 0.211 & 0.072 & 0.040 & 0.162\\
        Qwen2.5-72B & 0.157 & 0.333 & 0.289 & 0.124 & 0.088 & 0.194 & 0.061 & 0.040 & 0.161\\
        DeepSeek-R1-Distill-Llama-70B & 0.221 & 0.188 & 0.218 & 0.035 & 0.138 & 0.185 & 0.041 & 0.025 & 0.131\\
        GPT-4o & 0.249 & 0.375 & 0.365 & 0.097 & 0.089 & 0.240 & 0.009 & 0.026 & 0.181\\
        \midrule
        Linear Probe Llama-3.2-11B & 0.279 & 0.373 & 0.300 & 0.140 & 0.202 & 0.235 & 0.048 & 0.043 & 0.202\\
        LLMAO-Llama-3.2-11B & 0.144 & 0.190 & 0.188 & 0.078 & 0.118 & 0.116 & 0.063 & \underline{0.113} & 0.126\\
        WELL-CodeBERT & 0.090 & \textbf{0.575} & \underline{0.532} & 0.129 & 0.094 & 0.111 & \textbf{0.216} & 0.059 & 0.226\\
        WELL-Llama-3.2-11B & 0.236 & 0.028 & 0.139 & 0.000 & 0.054 & 0.081 & 0.000 & 0.000 & 0.067\\
        GridLoc-Llama-3.2-11B & \underline{0.291} & 0.498 & 0.206 & \underline{0.332} & \textbf{0.262} & \textbf{0.339} & \underline{0.158} & 0.039 & \underline{0.266}\\
        \midrule
        \ourmethod{}-Llama-3.2-11B & \textbf{0.334} & \textbf{0.575} & \textbf{0.568} & \textbf{0.481} & \underline{0.237} & \underline{0.291} & 0.096 & \textbf{0.217} & \textbf{0.350}\\
        \bottomrule
        \end{tabular}
\end{table*}

\textbf{Defects4J}
Bugs along with commits to fix the bug.
We use two versions of the dataset: the standard Defects4J v1.2.0 \citep{defects4j} dataset containing 395 bugs, and the additional 543 bugs from Defects4J v3.0.1 released in November 2024 which we use as a stronger evaluation of generalization. In both versions, we split each buggy file into a set of buggy methods and evaluate on the method level following \citet{wu2023large}.

\textbf{GithHub-Python} \citep{yasunaga2021break} and \textbf{GithHub-Java} \citep{santos2018syntax} consist of code mined from GitHub with syntax errors in Python and Java respectively.

\textbf{DeepFix} Real C programs written by students, some containing beginner syntax mistakes \citep{gupta2017deepfix}.

\textbf{TSSB-3M} ``Simple, stupid bugs'' (SStuBs) in Python which are mined from GitHub and categorized \citep{richter2022tssb}. Despite their name, these bugs are extremely challenging for humans to localize.

\textbf{ManySStuBs4J} Java SStuBs \citep{karampatsis2020often}.

\textbf{Juliet-Java} \citep{juliet-java} and \textbf{Juliet-C} \citep{juliet-cpp} consist of synthetic code corresponding to Common Weakness Enumerations (CWEs). We rename variables and function names to remove indicators of the vulnerability. We also remove comments and rename imports that refer to the dataset names. In total, we consider 89 CWEs from both the Juliet datasets.

The datasets are summarized in Table~\ref{tab:datasets} and we provide a detailed breakdown of the datasets in Appendix~\ref{app:datasets}.


\subsection{Baselines}

We group the baselines into three types: traditional FL methods that require code execution, LLM prompting of different models, and LLM probing/training.

\textbf{Traditional FL Methods}
For methods requiring code execution, we compare with DeepFL \citep{deepfl}, SmartFL \citep{smartfl}, and TRANSFER-FL \citep{transferfl} on Defects4J since these are the best performing traditional FL methods. Results for DeepFL and Transfer-FL are from \citet{llmao}, and we cite SmartFL results directly from \citet{smartfl}. We can not evaluate these benchmarks on the other datasets since they do not provide tests.

\textbf{Prompting Methods}
For models, we use a diverse set of four open-weights LLMs of size $\sim$70B parameters as well as a proprietary model. We consider Llama 3.3 70B \citep{llama3}, Qwen 2.5 72B \citep{qwen2}, CodeLlama 70B \citep{codellama}, and DeepSeek-R1-Distill-Llama-70B \citep{r1} as the main open-weights LLMs and we use GPT-4o \citep{gpt4o} as the representative proprietary LLM for prompting experiments. Llama 3.3 70B and Qwen 2.5 72B are LLMs pretrained on a diverse dataset of natural language and code while CodeLlama 70B additionally trained on primarily code data \citep{codellama}. DeepSeek-R1-Distill-Llama-70B is a ``reasoning'' model in that it can use longer chain-of-thought output to solve more complex problems \citep{r1}.

We experimented with several prompts based on that used by \citet{wu2023large}. Our prompt
asks for the buggy line text as well as the line number in case the LLM cannot count lines. We use this prompt with temperature 0 sampling for all models.
The prompt is provided in Appendix~\ref{app:prompts}.

\textbf{Probing Methods}
Finally, for LLM probing, we consider the following baselines:
\begin{itemize}[itemsep=0.0pt, topsep=0pt]
    \item Linear Probing: performing logistic regression on the last token representation from each code sample to predict if the code is buggy or not. Following \citet{repe}, we apply the trained classifier to each token in the sequence for getting token-level scores and then we derive line rankings the same way as \ourmethod{}.
    \item GridLoc \citep{niu2022does}: the only other attention probing method we are aware of, which uses an RNN to learn attention weights. We apply this method to learning the bug detection task from the dataset $\mathcal{H}_\text{Det}$, and interpret the resulting attention weights for FL using our line-aggregation method.
    \item LLMAO \citep{llmao}: trains an adapter using strong FL supervision. This method was designed for the CodeGen \citep{nijkamp2023codegen} models, and we used the largest available CodeGen model. We additionally adapted their code for the Llama-3 family of models and report results on Llama-3.2-11B.
    \item WELL \citep{well}: the first method to consider learning FL without strong supervision. This method finetunes a bidirectional LLM (CodeBERT \citep{codebert}) for bug detection and then interprets the attention weights of the last layer for fault localization. We additionally report results on Llama-3.2-11B, even though it has exclusively causal attention.
\end{itemize}

\subsection{RQ1: FL Performance}

For FL, given a program fragment with bugs, each method ranks the lines of the input program based on the likeliness of the bug being located to that line. We note that our method was not trained with direct localization information and instead makes use of weak supervision even though baselines (such as LLMAO) use strong FL supervision. WELL is the only baseline which uses weak supervision, similar to our method \citep{well}.



FL performance of \ourmethod{} and baselines on eight datasets is shown in Table~\ref{tab:defects4j}. To compare with existing work on Defects4J, we perform 10-fold cross validation using method-level context following \citep{wu2023large}. Results for DeepFL, TRANSFER-FL, and Smart-FL are from \citet{llmao} and \citet{smartfl}. Note that ``\na'' results indicate baselines which could not be evaluated due to requiring tests which were only available for Defects4J.

\ourmethod{} improves over the strongest baseline by 34.6\% top-1 accuracy on average across all eight datasets, in particular improving
on Defects4J v1.2.0 by 19.7\% and on Juliet-C by 128\% on top-1 accuracy.
The second highest performing method is GridLoc, which is a probing method we adapted for this setting.

We find that training a classifier from weak supervision in the form of fault detection is not enough for strong fault localization. This is seen by the WELL baseline using Llama-3.2-11B performing worse than random guessing. We find that WELL with Llama-3.2-11B almost always predicts the first line of code, which rarely is buggy. This issue does not happen when using CodeBERT, a model with bidirectional attention.

We also observe that prompting LLMs of size $\sim$70B parameters outperforms traditional FL methods requiring code execution. These 70B LLMs have strong performance since the proprietary GPT-4o (rumored to be over 200B parameters) performs similar to Llama 3.3 70B. 
In addition, we find that DeepSeek-R1-Distill-Llama-70B performs significantly worse than Llama-3.3-70B, the model it was finetuned from. By observing many of the outputs from this model, the extensive chain-of-thought (over 1000 tokens on average) is overly cautious and concludes that even benign lines could be buggy.
We discuss the scaling of performance of prompting and probing with model size in Section~\ref{sec:efficiency}.

\subsection{RQ2: Efficiency}
\begin{table}[h!]
\centering
\small
\caption{Resource efficiency across methods for localizing Defects4J bugs. We measure this through GPU overhead cost in gigabytes and the expected inference cost in FLOPs. }
\label{tab:memory_overhead}
\begin{tabular}{@{}lrrr@{}}
\toprule
\textbf{Model} & \textbf{Top-1} & \textbf{GPU Cost (GB)} & \textbf{$\mathbb{E}$[FLOPs]} \\ \midrule
Llama-3.2-90B  & 0.269 & 170.0 & \num{3.4e13} \\
CodeLlama-70B & 0.212 & 131.7 & \num{7.9e12} \\
Qwen2.5-72B& 0.157 & 138.9 & \num{4.2e12} \\
Llama-3.3-70B  & 0.269 & 154.0 & \num{3.4e13} \\
DeepSeek-R1DL-70B & 0.221 & 141.2 & \num{1.4e14}\\
\midrule
\ourmethod{}-Llama-3.2-1B & \underline{0.282} & \textbf{6.2} &\textbf{\num{2.0e9}} \\
\ourmethod{}-Llama-3.2-11B & \textbf{0.334} & \underline{24.2} & \underline{\num{2.2e10}} \\ \bottomrule
\end{tabular}
\end{table}

\label{sec:efficiency}
\begin{figure}[t]
    \centering

    \includegraphics[width=\linewidth]{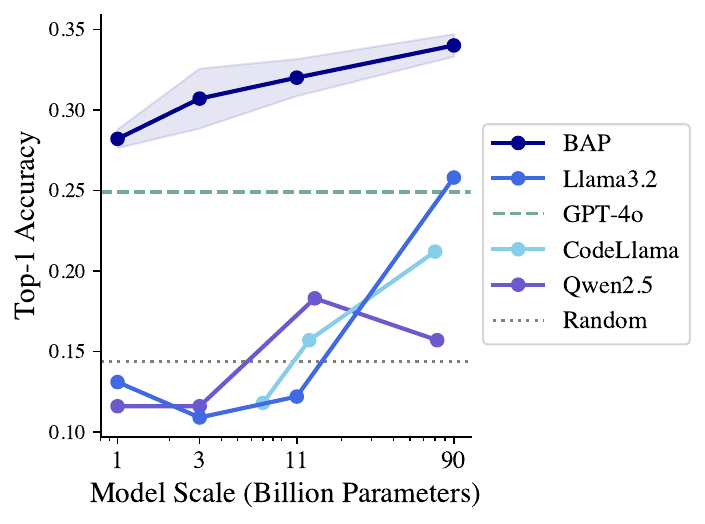}
    \caption{Model scale versus Top-1 on Defects4J. Each point for \ourmethod{} is trained on the hidden representations from the Llama-3.2 model of the corresponding size.}
    \label{fig:scale}
\end{figure}

We compare the size of the model which \ourmethod{} is trained on to the model's zero-shot fault localization capabilities in Figure~\ref{fig:scale}. We focus on top-1 accuracy for Defects4J since it is the most widely used benchmark for FL tools.
Figure~\ref{fig:scale} shows that \ourmethod{} trained via weak supervision on a 1B LLM outperforms zero-shot prompting of LLMs of size 70B or larger on Defects4J.
\ourmethod{} always outperforms its underlying model when zero-shot prompted across model sizes, but the performance gap shrinks as model size increases.

\ourmethod{} also displays a linear increase in performance with an exponential increase in model size while prompting results resemble ``emergent behavior'' \citep{wei2022emergent}.
To test the limits of \ourmethod{}s performance lead as model size continues to grow,
we scaled model size as far as practical with our resources, evaluated GPT-4o via API, and evaluated the DeepSeek-R1-Distill-Llama-70B model which leverages test-time scaling, but we never exceeded \ourmethod{}s performance through prompting. Surprisingly, the DeepSeek reasoning model even performed worse than Llama-3.3-70B, the model it was trained from.


\subsection{RQ3: Localizing Multi-line Bugs}
Since our method produces a ranking for every line of the input, our method is better suited for directly finding multiple bugs at once, or multi-line bugs. Since the top-k accuracy metric only cares if at least one of the top $k$ predictions are correct, we additionally use the precision at $k$ (P@$k$) metric which measures the percent of true buggy lines in the top $k$ predictions. The exact formula is provided in Appendix~\ref{app:precision}. We compare \ourmethod{} and baselines on Defects4J v1.2.0 in terms of P@$k$ in Table~\ref{tab:precision_k}.
\begin{table}[h]
    \centering
    \small
    \caption{Precision@K for multi-line bugs. We evaluate on a subset of Defects4J where there are strictly two or more buggy lines present within each function.}
    \begin{tabular}{lccc}
        \toprule
        \textbf{Method} & \textbf{P@2} & \textbf{P@3} & \textbf{P@5} \\
        \midrule
        Random & 0.201 & 0.231 & 0.297 \\
        \midrule
        CodeLlama-70B & \underline{0.250} & 0.284 & 0.351 \\
        Llama-3.3-70B & 0.240 & 0.266 & 0.355 \\
        Qwen2.5-72B & 0.221 & 0.271 & 0.347 \\
        DeepSeek-R1-Distill-Llama-70B & 0.245 & 0.283 & 0.336 \\  
        GPT-4o & 0.218 & \underline{0.288} & \underline{0.359} \\
        \midrule
        \textit{BAP}-Llama-3.2-11B & \textbf{0.289} & \textbf{0.298} & \textbf{0.367} \\
        \bottomrule
    \end{tabular}
    \label{tab:precision_k}
\end{table}

\subsection{RQ4: New Bug and Length Generalization}
Apart from efficiency as discussed above, we investigate several differences between \ourmethod{} and zero-shot prompting of the underlying model.


\textbf{New Bug Generalization}
We evaluate \ourmethod{} compared to LLM prompting and probing on 543 new bugs from Defects4J v3.0.1 in Table~\ref{tab:defects4j-3}. Our method and probing baselines are only trained on Defects4J v1.2.0, so this serves as an unbiased evaluation on the ability of these methods to generalize to new bugs. Our method, as well as the baselines, drop in performance for the new bugs, but \ourmethod{} maintains the highest top-$k$ accuracy (14.4\% increase in top-1 over the strongest baseline).

\begin{table}[t]
    \centering
    \small
    \caption{Comparison of \ourmethod{} with current state-of-the-art test-free fault localization method LLMAO on 437 additional bugs from Defects4J v3.0.1.}
    \label{tab:defects4j-3}
        \begin{tabular}{lrrr}
        \toprule
        & \multicolumn{3}{c}{\textbf{Defects4J v3.0.1}} \\
        \cmidrule(lr){2-4}
        \textbf{Method} & \textbf{Top-1} & \textbf{Top-3} & \textbf{Top-5} \\
        \midrule
        Random & 0.166 & 0.377 & 0.512\\
        \midrule
        CodeLlama-70B & 0.152 & 0.276 & 0.326 \\
        Llama-3.3-70B & \underline{0.215} & \underline{0.416} & 0.528 \\
        Qwen-2.5-72B & 0.161 & 0.395 & 0.515 \\
        GPT-4o & 0.181 & 0.451 & 0.579 \\
        \midrule
        Linear Probe Llama-3.2-11B & 0.156 & 0.409 & \underline{0.557} \\
        LLMAO-Llama-3.2-11B & 0.174 & 0.389 & 0.515 \\
        GridLoc-Llama-3.2-11B & 0.165 & 0.377 & 0.512 \\
        \midrule
        \ourmethod{}-Llama-3.2-11B & \textbf{0.246} & \textbf{0.459} & \textbf{0.583} \\
        \bottomrule
        \end{tabular}
\end{table}

\begin{figure*}[t]
    \centering
    \begin{subfigure}[b]{0.48\linewidth}
    \centering
        \includegraphics[width=\linewidth]{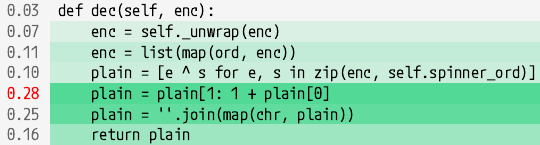}
        \caption{Python syntax error}
    \end{subfigure}
    \hfill
    \begin{subfigure}[b]{0.48\linewidth}
    \centering
        \includegraphics[width=\linewidth]{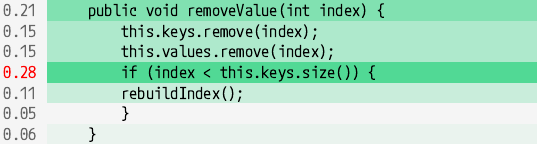}
        \caption{Defects4J bug}
    \end{subfigure}
    \caption{Examples of bug localization with \ourmethod{} on two evaluation set samples. We visualize the line-level weights from \ourmethod{} above such that lines highlighted in a darker color have higher weights. \ourmethod{} correctly identifies bug locations at Top-1. 
    }
    \label{fig:qualitative}
\end{figure*}

\textbf{Context Length Generalization}
\begin{figure}[t]
    \centering
    \includegraphics[width=\linewidth]{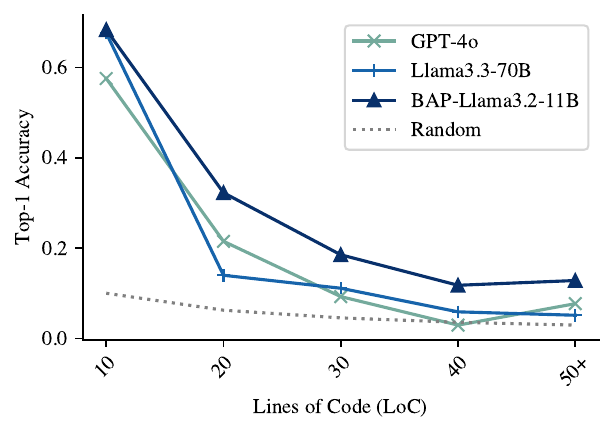}
    \caption{Top-1 accuracy versus context length, measured by lines of code (LOC) on Defects4J. We compare \ourmethod{}-Llama3.2-11B against models at least six times larger.}
    \label{fig:length-gen}
\end{figure}
We compare the behavior of \ourmethod{} to prompting in terms of length generalization in Figure~\ref{fig:length-gen}. We see that \ourmethod{} outperforms zero-shot prompting of various LLMs across context lengths from 10 to 50 lines. On code fragments of 60 lines and longer, all methods perform near random, making fault localization on code over 50 lines a challenging, but valuable task for future work.

We visualize the output of \ourmethod{} on two examples in Figure~\ref{fig:qualitative}.

\section{Related Work}

We survey related work in FL techniques and LLM probing.

\textbf{Automated Fault Localization.}
Methods for FL include the traditional spectrum-based (SBFL) and mutation-based (MBFL) methods which require executable code and deep-learning based approaches \citep{wong2016survey}. SBFL methods are simple but have low accuracy while MBFL and deep learning approaches have higher accuracy at larger computational cost \citep{wong2016survey}. Various deep learning approaches combine SBFL and MBFL with semantic features from deep models \citep{deepfl, transferfl, smartfl}. Recently, LLMs have significantly outperformed SBFL and MBFL approaches on FL on the method level \citep{wu2023large}. Prompting and agent-based systems can even perform repository-level FL \citep{agentless, agentfl}, but they must reduce the problem to method-level FL \citep{agentless}. LLMAO \citep{llmao} trains an adapter on an LLM from strong FL supervision to perform FL without executable tests, and WELL \citep{well} finetunes an LLM on bug detection supervision and interprets the attention for FL.
Unlike these approaches, our method uses LLM probing, and we leverage bug detection supervision to scale to more available data.

\textbf{Probing LLMs.}
Probing is useful tool in LLM interpretability.
There is extensive work on probing LLMs, most notably BERT \citep{devlin-etal-2019-bert}, to understand what linguistic knowledge it encodes. \citet{hewitt2019structural} design a probe for eliciting natural language syntax parse trees from BERT, and \citet{astprobe} probe for the code abstract syntax trees.
These probes are usually trained on a fixed size input \citep{repe}, but pooling sequence representations using global weights \citep{tenney2018you} and sample-conditional weights \citep{niu2022does} have been studied. Unlike these approaches, we adopt a traditional Transformer layer as our probe where the attention module learns to pool the input tokens.
\section{Conclusion}
In this paper, we approach the problem of scalable FL, and propose a method for achieving state-of-the-art FL performance at a fraction of the model inference cost by training on available data without strong FL supervision.
Existing methods for fault localization either require executable code, test cases, finegrained line-level supervision, or resource intensive LLMs.
To this end, we propose \ourmethodlong{} (\ourmethod{}), an LLM probing method which uses an attention mechanism to learn to localize bugs from only coarse-grained bug detection supervision.
Using a suite of eight diverse FL benchmarks, we demonstrate that \ourmethod{} significantly outperforms existing fault localization techniques as well as LLM prompting of models over ten times larger than that used by our probe.
We also identify avenues for future research including FL on long-code samples (over 50 lines), creation of more bug detection datasets by running existing bug detectors on large repositories of code, and execution-free FL on the file and project-level.

\section*{Acknowledgements}
This material is based upon work supported by the National
Science Foundation Graduate Research Fellowship under
Grant No. DGE-2236662 and the Google Research Fellowship.

\section*{Impact Statement}

This paper presents work whose goal is to advance the field of 
Machine Learning. There are many potential societal consequences 
of our work, none which we feel must be specifically highlighted here.

\bibliography{refs}
\bibliographystyle{iclr2025_conference}

\appendix
\newpage
\onecolumn

\section{Additional Experimental Details}
\label{app:experiment-details}

\subsection{Datasets}
\label{app:datasets}

The detailed breakdown of each of the datasets we used, other than Defects4J. We split datasets into groups based on the three domains of syntax, single line bugs, and vulnerabilities and show the breakdown in Table~\ref{tab:syntax-datasets}, \ref{tab:sstubs-datasets}, and \ref{tab:vuln-datasets} respectively.

\begin{table}[h]
    \centering
    \caption{Number of samples in syntax datasets.}
    \label{tab:syntax-datasets}
    \begin{tabular}{lrrrrrr}
    \toprule
         Subtype & \multicolumn{2}{c}{GitHub-Python} & \multicolumn{2}{c}{GitHub-Java} & \multicolumn{2}{c}{DeepFix}\\
         \cmidrule(lr){2-3}\cmidrule(lr){4-5}\cmidrule(lr){6-7}
         & Train & Test & Train & Test & Train & Test\\
         \midrule
         Correct syntax & 1323 & 400 & 1370 & 460 & 1475 & 365\\
         Mismatched parentheses & 400 & 100 & 110 & 100 & 400 & 100\\
         Mismatched bracket & 368 & 100 & 60 & 60 & 81 & 31\\
         Mismatched brace & 155 & 100 & 400 & 100 & 195 & 34 \\
         Missing semicolon & --- & --- & 400 & 100 & 400 & 100 \\
         Python-specific & 400 & 100 & --- & --- & --- & ---\\
         Java-specific & --- & --- & 400 & 100 & --- & --- \\
         C-specific & --- & --- & --- & --- & 400 & 100 \\
         \midrule
         Total & 2646 & 800 & 2740 & 920 & 2950 & 730\\
    \bottomrule
    \end{tabular}
\end{table}

\begin{table}[h]
    \centering
    \caption{Number of samples in SStuBs datasets.}
    \label{tab:sstubs-datasets}
    \begin{tabular}{lrrrr}
    \toprule
         Subtype & \multicolumn{2}{c}{TSSB} & \multicolumn{2}{c}{ManySStuBs}\\
         \cmidrule(lr){2-3}\cmidrule(lr){4-5}
         & Train & Test & Train & Test\\
         \midrule

No bug & 3745 & 1080 & 3821 & 1093\\
Change identifier & 400 & 100 & 400 & 100\\
Change numeral & 400 & 100 & 400 & 100\\
Change binary operator & 400 & 100 & 400 & 100\\
Change unary operator & 202 & 100 & 301 & 100\\
Less specific if & 281 & 100 & 400 & 100\\
More specific if & 400 & 100 & 400 & 100\\
Same function less args & 400 & 100 & 266 & 100\\
Same function more args & 400 & 100 & 400 & 100\\
Swap arguments & 81 & 80 & 94 & 93\\
Swap boolean literal & 381 & 100 & 360 & 100\\
Python specific & 400 & 100 & --- & ---\\
Java specific & --- & --- & 400 & 100\\
\midrule
Total & 7830 & 2184 & 7642 & 2186\\
    \bottomrule
    \end{tabular}
\end{table}

\begin{table}[h]
    \centering
    \caption{Number of samples in security vulnerabilities datasets.}
    \label{tab:vuln-datasets}
    \begin{tabular}{lrrrr}
    \toprule
         CWE Class & \multicolumn{2}{c}{Juliet-Java} & \multicolumn{2}{c}{Juliet-C}\\
         \cmidrule(lr){2-3}\cmidrule(lr){4-5}
         & Train & Test & Train & Test\\
         \midrule
         Access Control & 677 & 178 & 609 & 159 \\
Comparison & 38 & 11 & 14 & 4 \\
Concurrency & 21 & 11 & 172 & 48 \\
Encryption & 616 & 154 & 278 & 72 \\
Exposed Resource & 574 & 141 & 934 & 236 \\
File Handling & 1056 & 256 & --- & --- \\
Implementation & 104 & 27 & 144 & 37 \\
Improper Check or Handling of Exceptional Conditions & 37 & 10 & 604 & 152 \\
Improper Input Validation & 523 & 128 & 304 & 76 \\
Improper Neutralization & --- & ---  & 60 & 16 \\
Incorrect Calculation & 40 & 10 & 99 & 27 \\
Injection & 2368 & 571 & 304 & 76 \\
Insufficient Control Flow Management & 161 & 49 & 259 & 69 \\
Memory Safety & 342 & 86 & 847 & 217 \\
Poor Coding Practices & 665 & 175 & 1820 & 465 \\
Protection Mechanism Failure & --- & --- & 28 & 8 \\
Randomness & 52 & 14 & 14 & 4 \\
Resource Control & 534 & 126  & --- & --- \\
Resource Lifecycle Management & 80 & 22 & 841 & 215 \\
Sensitive Information Exposure & 112 & 31 & 84 & 24 \\
    \bottomrule
    \end{tabular}
\end{table}

\subsection{Few-shot Prompts}
\label{app:prompts}
For all prompting experiments, we use zero-shot prompting with the prompt given below:

\begin{lstlisting}[frame=single]
Q: Please analyze the following code snippet for potential bugs. Return the fault
localization result in JSON format, consisting of five JSON objects called
"faultLocalization". These "faultLocalization" objects correspond to the top
five most suspicious lines of code. Each "faultLocalization" contains two
fields: 1) "codeContent" string which contains the line of code that corresponds
to suspicious code in the snippet and 2) "lineNumber" integer which indicates the
line number of this suspicious code. Output just the JSON objects
"faultLocalization" and NOTHING ELSE.
```
{code-example}
```
\end{lstlisting}





\subsection{Hyperparameters}
\label{app:hyperparams}

For \ourmethod{}, we trained for 30 epochs with a learning rate of \num{1e-4}, a batch size of 16, and weight decay of 1 for all datasets except for the TSSB dataset where we needed to use less training epochs to avoid overfitting. For TSSB, we trained for 5 epochs with a learning rate of \num{1e-4}, a batch size of 16, and weight decay of 1. For the architecture of \ourmethod{}, we used grouped query attentino with 32 query heads and 8 key-value heads to match the architecture of the Llama-3.2-11B attention mechanism. For the ablation of Llama-3.2-90B, we used 64 query heads with 8 key-value heads.

For the linear probing baseline and GridLoc, we trained for 30 epochs with a learning rate of \num{1e-4}, a batch size of 16, and weight decay of 0.1 for all datasets.


Parameters were chosen by splitting the training set with an 80/20 split into train and validation samples, and selecting hyperparameters from the results on the validation set.

\subsection{Compute resources}
\label{app:compute}

All experiments are conducted on a server with 96 Intel Xeon Gold 6248R CPUs, each with a clock speed of 3.00 GHz, and 8 NVIDIA A100 GPUs, each with a capacity of 40GB.

The WELL baseline is the most compute intensive of the methods we explore because it requires fintuning an LLM.
To finetune Llama-3.2-11B, we had to use LoRA \citep{hu2021lora} with rank 16 to make training this model accessible. GridLoc takes around twice the training time as \ourmethod{}, and all the other baselines take 2-3 minutes for one training run on a single dataset.

\section{Additional Results}\label{app:results}
\subsection{Error Bars for FL Results}
Error bars for the results in Table~\ref{tab:defects4j} are provided in Table~\ref{tab:defects4j-errors}. The prompting methods have no error bars because we use greedy decoding.

\begin{table*}[t]
    \centering
    \tiny
    \caption{Comparison of \ourmethod{} with existing fault localization methods across eight diverse bug benchmarks. We evaluate each method on line-level localization performance at the method-level, measured by top-1 localization accuracy. From left-to-right: Defects4J v1.2.0, GitHub-Python, GitHub-Java, DeepFix, TSSB-3M, ManySStuBs4J, Juliet-Java, and Juliet-C.}
    \label{tab:defects4j-errors}
        \begin{tabular}{lrrrrrrrr}
        \toprule
        \textbf{Method} & \textbf{D4J} & \textbf{GH-Py} & \textbf{GH-J} & \textbf{DeepFix} & \textbf{TSSB} & \textbf{MS4J} & \textbf{Juliet-J} & \textbf{Juliet-C} \\
        \midrule
        Random & 0.144 & 0.100 & 0.134 & 0.038 & 0.069 & 0.124 & 0.025 & 0.058\\
        \midrule
        DeepFL & 0.144 & \na & \na & \na & \na & \na & \na & \na\\      
        SmartFL & 0.158 & \na & \na & \na & \na & \na & \na & \na\\
        TRANSFER-FL & 0.218 & \na & \na & \na & \na & \na & \na & \na\\
        \midrule
        CodeLlama-70B & 0.212 & 0.145 & 0.316 & 0.084 & 0.077 & 0.169 & 0.038 & \underline{0.095}\\
        Llama-3.3-70B & 0.269 & 0.225 & 0.272 & 0.092 & 0.114 & 0.211 & 0.072 & 0.040\\
        Qwen2.5-72B & 0.157 & 0.333 & 0.289 & 0.124 & 0.088 & 0.194 & 0.061 & 0.040\\
        DeepSeek-R1-Distill-Llama-70B & 0.221 & 0.188 & 0.218 & 0.035 & 0.138 & 0.185 & 0.041 & 0.025\\
        GPT-4o & 0.249 & 0.375 & 0.365 & 0.097 & 0.089 & 0.240 & 0.009 & 0.026\\
        \midrule
        Linear Probe Llama-3.2-11B & 0.279$\pm$0.02 & 0.373$\pm$0.01 & 0.300$\pm$0.01 & 0.140$\pm$0.01 & 0.202$\pm$0.01 & 0.235$\pm$0.01 & 0.048$\pm$0.00 & 0.043$\pm$0.01\\
        LLMAO-CodeGen & 0.223 & \na & \na & \na & \na & \na & \na & \na\\
        LLMAO-Llama-3.2-11B & 0.144 & 0.190 &  & 0.078 &  &  &  & \\
        WELL-CodeBERT & 0.090 & \textbf{0.575} & \underline{0.532} & 0.129 & 0.094 & 0.111 & \textbf{0.216} & 0.059\\
        WELL-Llama-3.2-11B & 0.236 & 0.028 & 0.139 & 0.000 & 0.054 & 0.081 & 0.000 & 0.000\\
        GridLoc-Llama-3.2-11B & \underline{0.291$\pm$0.02} & 0.498$\pm$0.08 & 0.206$\pm$0.08 & \underline{0.332$\pm$0.03} & \textbf{0.262$\pm$0.03} & \textbf{0.339$\pm$0.03} & \underline{0.158$\pm$0.04} & 0.039$\pm$0.01\\
        \midrule
        \ourmethod{}-Llama-3.2-11B & \textbf{0.334$\pm$0.02} & \textbf{0.575$\pm$0.02} & \textbf{0.568$\pm$0.01} & \textbf{0.481$\pm$0.04} & \underline{0.237$\pm$0.02} & \underline{0.291$\pm$0.04} & 0.096$\pm$0.03 & \textbf{0.217$\pm$0.00}\\
        \bottomrule
        \end{tabular}
\end{table*}




\section{Precision at $k$}\label{app:precision}
The precision at $k$ (P@$k$) metric which we use is calculated as:
\begin{align*}
    \frac{\text{Correct in top $k$}}{\min(k, \text{Max possible correct})}.
\end{align*}
We use the min in the denominator to account for the case where the number of buggy lines is much fewer than $k$. This is practically relevant since many bugs consist of only 2-3 buggy lines which is less than $k$ for P@5.

\end{document}